\documentclass[12pt]{article}
\usepackage{amsmath}
\makeatletter
\renewcommand{\section}{\setcounter{equation}{0}\@startsection
  {section}%
  {1}%
  {0pt}%
  {-1\baselineskip}%
  {0.4\baselineskip}%
  {\large \bfseries}}%
\renewcommand{\subsection}{\@startsection
  {subsection}%
  {2}%
  {0pt}%
  {-0.75\baselineskip}%
  {0.2\baselineskip}%
  {\bfseries}}%
\renewcommand{\subsubsection}{\@startsection
  {subsubsection}%
  {3}%
  {0pt}%
  {-0.5\baselineskip}%
  {0.1\baselineskip}%
  {\sc}}%
\makeatother

  {\end{list}}%

\def\cJ{{\cal J}}

\def\A{{\rm A}}
\def\F{{\rm F}}
\def\T{{\rm T}}
\def\Taone{\T^{a_1}}
\def\Tatwo{\T^{a_2}}
\def\Tathree{\T^{a_3}}
 
\def\W{{\rm W}}
\def\Z{{\rm Z}}

\def\gambar{\bar\gamma}  
\def\teta{{1\over 2}\theta}
\def\iteta{{i\over 2}\theta}
\def\eps{\rm eps}

\def\Aslash{{\A\mkern-12mu/}}
\def\Dirac{{D\mkern-12mu/}}
\def\prslash{{\partial\mkern-9mu/}}
\def\poneslash{{p\mkern-8mu/}_1{\!}} 
\def\ptwoslash{{p\mkern-8mu/}_2{\!}}

\def\prslash{{\partial\mkern-9mu/}}    %%_standard_Dirac_operator
\def\qslash{{q\mkern-8mu/}{\!}}

\def\idx{\int\! d^4\!x \,}
\def\idp{\int\! \frac{d^4\!p}{(2\pi)^4} \,\,}
\def\idq{\int\! \frac{d^4\!q}{(2\pi)^4} \,\,}
\def\ds{\displaystyle}
\def\RR{{\rm I\!\!\, R}}

\begin{document}
\begin{titlepage}
\rightline{FT/UCM-10-2001}

\vskip 1.5 true cm
\begin{center}
{\Large \bf Chiral Gauge Anomalies on Noncommutative Minkowski Space-time}
\footnote{Written version of the invited lecture given at the 
Euroconference ``Brane New World and Noncommutative Geometry", Villa Gualino, 
 Torino, Italy, October 2-7 2000. To be published in the conference 
  proceedings by World Scientific.}\\ 
\vskip 1.2 true cm 
{\rm C.P. Mart\'{\i}n}
\vskip 0.3 true cm
{\it Departamento de F\'{\i}sica Te\'orica I}\\
{\it Facultad de Ciencias F\'{\i}sicas}\\ 
{\it Universidad Complutense de Madrid}\\
{\it 28040 Madrid, Spain}\\
{\rm E-mail}: {\tt carmelo@elbereth.fis.ucm.es}\\
\vskip 1.2 true cm

{\leftskip=45pt \rightskip=45pt 
\noindent
Chiral gauge anomalies on noncommutative Minkowski Space-time are computed and
have their origin elucidated. The consistent form and the covariant form of the anomaly are obtained. Both Fujikawa's method and Feynman diagram 
techniques are used to carry out the calculations. \par}

\end{center}

\end{titlepage}
\setcounter{page}{2}

%----------------------------------------------------- Paper

\section{Introduction}

 The study of chiral anomalies in field theories over commutative 
manifolds is a well-established subject  whose importance both in Physics and 
Mathematics cannot be overstated --see ref.~\cite{qftanom} and 
references therein. The concept of Noncommutative manifold~\cite{booka, bookb}
and the ideas and tools furnished by field theories defined on 
them may prove decisive~\cite{booka} in the quest of understanding the 
fundamental properties of Nature  --see ref.~\cite{WS} and 
references therein. It is therefore no futile endeavor to study the quantum 
properties of field theories over noncommutative manifolds which have chiral 
symmetries at the classical level. We shall be modest in this article  
and restrict our analysis to  field theories over noncommutative 
Minkowski Space-time. We shall study the quantization of  chiral fermions 
coupled to a classical background $U(N)$ gauge field when the space-time 
manifold is noncommutative Minkowski. We refer the reader to 
ref.~\cite{KonSch} for a quick introduction to noncommutative manifolds 
and, in particular, to  noncommutative space-time.

This article is divided into two main parts. In the first part we shall use
path integral methods --i.e. variants of Fujikawa's method-- to compute
the chiral gauge anomaly. We shall obtain thus the consistent form and the
covariant form of the gauge anomaly. We shall also see that in both cases 
the gauge anomaly occurs because of the lack of chiral gauge invariance of 
the fermionic measure. In the second part we shall understand
the UV and IR origin of the chiral anomaly by using diagrammatic techniques.
In this second part we shall compute the noncommutative counterpart of the
famous triangle anomaly for the gauge currents. Before we start off, let us 
warn the reader that no global anomaly will be considered in the sequel.

\section{The gauge anomaly in the path integral formalism}

Let us take a classical chiral, say, right handed, fermion, 
$\psi^j_{\rm R}\!=\!{\rm P}_{+}\psi^{j}$, (${\rm P}_{+}\!=\!\frac{1+\gamma_5}{2}$),  
carrying the fundamental representation of the group $U(N)$.
The physics of this fermion interacting 
with a $U(N)$ classical gauge field on noncommutative Minkowski space-time 
is given by the classical action 
\begin{equation}
{\rm S}\;=\;\idx \bar{\psi}_{i}\star(i\prslash\psi^{i} +
\A^{i}_{\mu\,j}\star\gamma^{\mu} {\rm P}_{+}\psi^{j}),
\label{chiralclassact}
\end{equation}
where $\psi^j$ is a Dirac fermion carrying the fundamental representation of
$U(N)$ and the complex matrix $\A^{i}_{\mu\,j}$, with 
$(\A^{i}_{\mu\,j})^*=\A^{j}_{\mu\,i}$, is the $U(N)$ gauge field. 
The indices $i,j$ run from $1$ to $N$. The previous action is invariant under
the following chiral gauge transformations: 
\begin{equation} 
\begin{array}{l}
{ \big(\delta_{\omega} \A_{\mu}\big)^i_{\;j} = 
\partial_{\mu}\omega^i_{\;j}-i\,\A^i_{\mu\,k}\star\omega^k_{\;j}+
i\,\omega^i_{\;k}\star\A^k_{\mu\,j}, } \\[12pt]
{(\delta_{\omega}\psi)^i=i\,\omega^i_{\;j}\star {\rm P}_{+}\,\psi^j\qquad{\rm and}
\qquad
(\delta_{\omega} \bar{\psi})_{k}=
-i\bar{\psi}_k\star\omega^{k}_{\;i}{\rm P}_{-},}   
\end{array}
\label{gaugetrans}
\end{equation} 
where ${\rm P}_{-}\!=\!\frac{1-\gamma_5}{2}$. The complex functions  
$\omega^i_{\;j}=\omega^{*\,j}_{\quad i}$, $i,j=1,\cdots, N$, 
are the infinitesimal gauge  transformation parameters. The symbol $\star$ 
denotes  the Moyal product of functions on Minkowski space-time.
The Moyal product is defined, in terms of Fourier transforms, thus
\begin{displaymath}
(f\star g)(x)= 
\idp\idq\;e^{i(p+q)_{\mu}x^{\mu}} \;
e^{-\frac{i}{2}\theta^{\mu\nu}p_{\mu}q_{\nu}}\;\hat{f}(p)\hat{g}(q).
\end{displaymath}
Here, $\theta^{\mu\nu}$ is an antisymmetric real matrix  either of  
magnetic type or light-like type. It is for these choices of 
matrix $\theta$ that it is likely that a unitary theory exists at the quantum 
level, this theory being connected by Wick rotation with its counterpart 
in noncommutative Euclidean space~\cite{AGM}.

The quantum theory in a background classical gauge field is given by
$\W[\A]$, this functional being formally defined as follows
\begin{equation}
 e^{-\W[\A]}\;\equiv\;\int d\psi\, d\bar{\psi}\;e^{-{\rm S}[\A]}\;\equiv\;\Z[\A].
\label{effective}
\end{equation}
To render the path integral above as well-defined as possible we are setting
our theory on noncommutative space-time with Euclidean signature. 
${\rm S}[\A]$ is therefore the Euclidean counterpart of the action in 
eq.~(\ref{chiralclassact}):
\begin{equation}
{\rm S}\;=\;\idx\;\bar{\psi}_i\, i\hat{D}(\A)^i_{\;\;j}\,\psi^j.
\label{euclidact}
\end{equation}
The operator $i\hat{D}(\A)$, which acts on Dirac spinors $\psi^j$ as follows 
\begin{equation}
 i\hat{D}(\A)^{i}_{\;\;j}\,\psi^j =i\prslash\psi^{i} +
\A^{i}_{\mu\,j}\star\gamma^{\mu} {\rm P}_{+}\psi^{j},
\label{chiralop}
\end{equation}
is not an hermitian operator, but it has a well-defined eigenvalue 
problem since it is an elliptic operator. Actually, the principal symbol 
of this operator is the same as that of its counterpart in ordinary  
(commutative) space.  Note that we take $\gamma^\mu=(\gamma^\mu)^{\dagger}$ and
$\gamma_5=-\gamma^1\gamma^2\gamma^3\gamma^4$.
 
Now, for the theory to make sense at the quantum level, we must demand that
$\W[\A]$ in eq.~(\ref{effective}) be invariant under the gauge transformations
in eq.~(\ref{gaugetrans}), i.e. 
\begin{equation}
\delta_{\omega}\W\,=\,0\quad \Longleftrightarrow\quad
\idx\,\omega^i_{\;j}(x)\, \big(D_{\mu}[\A]\,\cJ_\mu\big)^{j}_{\;\;i}(x)\,=\,0.
\label{gaugeinvofW}
\end{equation}
Here, $D_{\mu}[\A]^{i}_{\;\;j}(\cdot)\,=\,\partial_{\mu}\delta^{i}_{\;\;j}\,
(\cdot)
- i\,\A^i_{\mu\,j}\star\,(\cdot) +
i\,(\cdot)\star\A^i_{\mu\,j}$ is the covariant derivative. The current
$\cJ^{i}_{\mu\,j}$ is formally defined as follows
\begin{equation}
\cJ^{i}_{\mu\,j}(x)\,\equiv\, 
\langle \big(\psi^j_{\beta}\star\bar{\psi}_{\alpha\,i}\big)(x)
\big(\gamma_{\mu}{\rm P}_{+}\big)_{\alpha\beta}\rangle
\,=\,\frac{\delta \W[\A]}{\delta\A^j_{\mu\,i}(x)},
\label{formalcurrent}
\end{equation}
where 
\begin{equation}
\langle\cdots\rangle\,=\,
\frac{\int d\psi d\bar{\psi}\;\cdots\;e^{-{\rm S}[\A]}}
{\int d\psi d\bar{\psi}\;e^{-{\rm S}[\A]}}.
\label{VEV}
\end{equation}

We shall see below that eq.~(\ref{gaugeinvofW}) does not hold for the quantum
theory in spite of the fact that the classical theory is gauge invariant: 
the gauge symmetry is anomalous at the quantum level.
To make correct statements on the properties of $\W[\A]$, one must properly 
define first the formal path integral in eq.~(\ref{effective}). Since the 
fermion fields are Grassman fields and the classical action in 
eq.(\ref{euclidact})  
is quadratic in those fields, one aims to define, upon renormalization,  
$\Z[\A]$  as the determinant of a certain operator with 
eigenvalues, say,  $\big\{\lambda_n\big\}_n$:
\begin{displaymath}
 \Z[\A]\;=\;\Big[\prod_{n}\lambda_n\Big]_{\rm ren}. 
\end{displaymath}
As in ordinary (commutative) Euclidean space there are two conspicuous choices
of $\big\{\lambda_n\big\}_n$, namely:
\begin{description}
\item[a)] $\big\{\lambda_n\big\}_n$ is the set of eigenvalues of 
$i\hat{D}(\A)$.
\item[b)] $\big\{\lambda_n\big\}_n$ is the set of eigenvalues of 
$\sqrt{\Big(i\hat{D}(\A)\Big)^{\dagger}i\hat{D}(\A)}$.
\end{description}
The operator $i\hat{D}(\A)$ has been defined in eq.(\ref{chiralop}) and 
$\Big(i\hat{D}(\A)\Big)^{\dagger}$ is its hermitian conjugate. We shall 
show shortly that choices $a)$ and $b)$ lead, respectively, to the consistent 
form and the covariant form of the anomaly. The consistent form --unlike 
the covariant form-- satisfies 
the noncommutative Wess-Zumino condition~\cite{WZ} and the covariant 
form transforms covariantly under gauge transformations. The consistent form 
does not transform  covariantly under gauge transformations. As in the 
ordinary (commutative) Euclidean space case, for choice b), the current 
$\cJ^{i}_{\mu\,j}(x)\,=\, 
\langle \big(\psi^j_{\beta}\star\bar{\psi}_{\alpha\,i}\big)(x)
\big(\gamma_{\mu}{\rm P}_{+}\big)_{\alpha\beta}\rangle$ cannot be obtained, 
upon renormalization, by functional differentiation of the renormalized
$\W[\A]$. Hence, the Quantum Action Principle does not hold in the theory 
constructed with choice b)(see ref.~\cite{QAP}, and references therein, for  
background information on the Quantum Action Principle). The reader may notice 
that the results we have just stated are already true if Euclidean space 
were commutative (see ref.~\cite{AGP} for case a) and ref.~\cite{Fuji} for 
case b), see also ref.~\cite{BarZu}).

\subsection{The consistent form of the anomaly}

We have seen above that $i\hat{D}(\A)$, defined in eq.(\ref{chiralop}), is an 
elliptic operator. Let us denote the set of its  eigenvalues by 
$\big\{\lambda_n\big\}_n$. Then, following ref.~\cite{AGP}, we introduce  
two basis of eigenspinors, say, $\{\phi_n\}_n$ and $\{\chi_n\}_n$ 
defined as follows
\begin{equation}
i\hat{D}(\A)\phi_n\,=\,\lambda_n\,\phi_n,\quad 
\Big(i\hat{D}(\A)\Big)^{\dagger}\chi_n\,=\,\lambda_n^{*}\,\chi_n,
\label{eigenspinors}
\end{equation}
where $\lambda_n^{*}$ is the complex conjugate of $\lambda_n$. We shall also 
normalize the eigenspinors so that the following equation holds
\begin{equation}
\idx \chi^{\dagger}_n(x)\phi_m(x)\;=\;\delta_{nm}.
\label{ortoga}
\end{equation}
Now, the Grassman spinor fields $\psi(x)$ and $\bar{\psi}(x)$ can be expanded
thus
\begin{equation}
\psi(x)\,=\,\sum_n\,a_n\,\phi_n(x),\quad
\bar{\psi}(x)\,=\,\sum_n\,\bar{b}_n\,\chi_n^{\dagger}(x).
\label{expansion}
\end{equation}
The coefficients $\{a_n\}_n$ and $\{\bar{b}_n\}_n$ being Grassman numbers.

To define the path integral in eq.~(\ref{effective}) we shall define its
measure first. We define
\begin{equation}
 d\psi\, d\bar{\psi}\equiv \prod_n\;da_n\,d\bar{b}_n.
\label{measure}
\end{equation}
Taking into account eq.~(\ref{euclidact}) and eqs.~(\ref{eigenspinors}) -- 
(\ref{expansion}), one readily shows that
\begin{displaymath}
{\rm S}\;=\;\sum_n\;\lambda_n\,\bar{b}_n\,a_n.
\end{displaymath}
Furnished with this equation and eq.~(\ref{measure}) we are ready to define
the path integral in eq.~(\ref{effective}) as follows
\begin{equation}
\int d\psi\, d\bar{\psi}\;e^{-{\rm S}[\A]}\;=\;
\int \prod_n\,  da_n\,d\bar{b}_n\;e^{-\sum_n\;\lambda_n\,\bar{b}_n\,a_n}.
\label{pathintegrala}
\end{equation}
One can now  work out  the Grassman integrations to obtain
\begin{displaymath}
\Z[\A]\;=\;\prod_n\,\lambda_n\;=\;{\mbox ``}{\rm det}\;i\hat{D}(\A)
{\mbox '}{\mbox '}.
\end{displaymath}   
Of course, the product of eigenvalues above is ill-defined and needs regularization. One can use Pauli-Villars (PV) regularization, or zeta function 
regularization, to define a regularized $\Z[\A]$ and then renormalized it:
\begin{equation}
\Z[\A]_{\rm PV\, regularized}\,=\,\prod_n\,\frac{\lambda_n}{\lambda_n+M}
\quad\stackrel{\rm renormalization}{\longrightarrow}\quad \Z[\A]\,=\,
\big({\rm det}\;i\hat{D}(\A)\big)_{\rm renormalized}.
\label{renorproc}
\end{equation} 
We shall not delve more deeply into the renormalization process since we 
will not need it to obtain the anomaly. Notice that $\Z[\A]$ in 
eq.~(\ref{renorproc}) cannot be gauge invariant since, under the 
infinitesimal gauge transformation, $g$, in  eq.~(\ref{gaugetrans}), the 
operator $i\hat{D}(\A)$ does not undergo a similarity transformation; rather, 
it changes as follows
\begin{displaymath}
(1-i\,\omega\,P_{-})\,i\hat{D}(\A^g)\,(1+i\,\omega\,P_{+})\,=\,i
\hat{D}(\A)\,+\,O(\omega^2).
\end{displaymath}
However, as also happens in the commutative space-time setting, the modulus
of the quantity  $\big({\rm det}\,i\hat{D}(\A)\big)_{\rm renormalized}$ 
can be defined in a gauge invariant gauge, and hence, it is the {\it phase} of 
$\big({\rm det}\,i\hat{D}(\A)\big)_{\rm renormalized}$ which carries the
anomaly. Indeed,
\begin{displaymath}
\mid{\rm det}\;i\hat{D}(\A)\mid^2\,=\,
{\rm det}\;(i\prslash_{+}i\prslash_{-})\,\big({\rm det}\;i\Dirac(\A)\big)^2,
\end{displaymath}
and the Dirac operator $i\Dirac (\A)$ can be regularized in a gauge invariant
way by using, for instance, the Pauli-Villars method. Note that
$\prslash_{\pm}\,=\,\prslash P_{\pm}$ and $i\Dirac (\A)=i\prslash+\Aslash$.
 
We now turn to the  computation of  the consistent form of the anomaly. 
Let us introduce the following definitions
\begin{equation}
\begin{array}{l}
\Z[\A+\delta_\omega \A]\,=\,
\int d\psi'\, d\bar{\psi}'\;e^{-{\rm S}[\A+\delta_{\omega}\A]}\;=\;
\int \prod_n\, da'_n\,d\bar{b}'_n\;
e^{-\sum_n\;\lambda_n[\A+\delta_{\omega}\A]\,\bar{b}'_n\,a'_n},\\
\Z[\A]\,=\,
\int d\psi\, d\bar{\psi}\;e^{-{\rm S}[\A]}\;=\;
\int \prod_n \, da_n\,d\bar{b}_n\;
e^{-\sum_n\;\lambda_n[\A]\,\bar{b}_n\,a_n},
\end{array}
\label{primes}
\end{equation}
where
\begin{displaymath}
\begin{array}{l}
\psi'\,=\,\psi\,+\,\delta_{\omega}\psi\,=\,\sum_n\,a'_n\,\phi_n,\quad
\bar{\psi}'\,=\,\bar{\psi}\,+\,\delta_{\omega}\bar{\psi}\,=\,\sum_n\,\bar{b}'_n\,\chi_n^{\dagger},\\
\psi\,=\,\sum_n\,a_n\,\phi_n,\bar{\psi}\,=
\,\sum_n\,\bar{b}_n\,\chi_n^{\dagger}, 
\end{array}
\end{displaymath}
and $\delta_{\omega}\A$, $\delta_{\omega}\psi$ and $ \delta_{\omega}\bar{\psi}$
are given in eq.~(\ref{gaugetrans}), and $\phi_n$ and $\chi_n$ are defined
in eq.~(\ref{eigenspinors}). Next, let us define $\delta\,J$ as follows:
\begin{equation}
\delta\,J\,\equiv\,-i\idx\sum_{n}\{\chi^{\dagger}_n
\star\omega\star P_{+}\phi_n-
\chi^{\dagger}_n\star\omega\star P_{-}\phi_n\}\,=\,
-i\idx\sum_{n}\{\chi_n^{\dagger}\star\omega\star \gamma_5\phi_n\}.
\label{deltajacobian}
\end{equation}
Then, it is not difficult to show that, under the gauge transformations in  
eq.~(\ref{gaugetrans}), the measure of  the path integrals in 
eq.~(\ref{primes}) changes  thus
\begin{equation}
\prod_n  da'_n\,d\bar{b}'_n\,-\,\prod_n  da_n\,d\bar{b}_n\,=\,
\delta J\,
\prod_n  da_n\,d\bar{b}_n\,+\,O(\omega^2).
\label{measurevar}
\end{equation}
Finally, taking into account that 
$\delta_{\omega}\Z[\A]\,=\,\Z[\A+\delta\A]-\Z[\A]\,+\,
O(\omega^2)$ and eqs.~(\ref{primes}) -- (\ref{measurevar}), one obtains
that 
\begin{equation}
\delta_\omega\W[\A]\,=\,-\idx\,\omega^i_{\;j}(x)\, \big(D_{\mu}[\A]\,\cJ^{\rm consistent}_\mu\big)^{j}_{\;\;i}(x)\,=\,-
i\,\idx\sum_{n}\{\chi_n^{\dagger}\star\omega\star \gamma_5\phi_n\}.
\label{formalanom}
\end{equation}
The current, $\cJ^{\rm consistent}_\mu$, is the current, 
$\cJ_\mu$, in eq.~(\ref{formalcurrent}), when the path 
integrals in eq.~(\ref{VEV}) are defined as in eq.~(\ref{pathintegrala}).
The current has been labeled ``consistent'' since as we shall see, when 
properly defined, its gauge divergence  satisfies  the Wess-Zumino 
consistency condition. 
The right hand side of eq.~(\ref{formalanom}) needs regularization. 
To regulate it we shall use the Gaussian cut-off function furnished by 
$\{\lambda_n^2\}_n$:
\begin{equation}
\idx\,\omega^i_{\;j}(x)\, \big(D_{\mu}[\A]\,\cJ_\mu^{\rm consistent}
\big)^{j}_{\;\;i}(x)\,=\,i\,\lim_{\Lambda\rightarrow\infty}\idx\sum_{n}
\{\chi_n^{\dagger}\star\omega\star \gamma_5\, 
e^{-\frac{\lambda^2_n}{\Lambda^2}}\,\phi_n\}.
\label{cutoffed} 
\end{equation} 
By changing to a plane wave basis,  the right hand side of 
eq.~(\ref{cutoffed}) can be recast thus
\begin{equation}
{\ds i\,\lim_{\Lambda\rightarrow\infty}\idx \omega^{i}_{\;\,j}(x)\,
{\rm tr}\,\{\gamma_5\,\big(e^{\frac{\hat{D}^2}{\Lambda^2}}e^{ipx}\big)^{j}_{\;\,i}
\star e^{-ipx}\}.}
\label{plane} 
\end{equation} 
Here, ``${\rm tr}$'' denotes trace over the
Dirac spinor indices. After a long computation~\cite{Pepe},  
eq.~(\ref{plane}) leads to
\begin{displaymath}
\idx\,\omega^i_{\;j}(x)\, 
\big(D_{\mu}[\A]\,\cJ_\mu^{\rm consistent}\big)^{j}_{\;\;i}(x)\,=\,
{\cal A}^{\rm consistent}(\A,\omega),
\end{displaymath}
with
\begin{equation} 
{\cal A}^{\rm consistent}(\A,\omega)\,=\,
-{i\over 24\pi^2}\,{\rm Tr}\int d^4 x\,
\varepsilon_{\mu_1\mu_2\mu_3\mu_4}\,\omega\,\partial_{\mu_1}\bigl[
\A_{\mu_2}\star\partial_{\mu_3}\A_{\mu_4}- {i\over
2}\A_{\mu_2}\star\A_{\mu_3}\star\A_{\mu_4}\bigr].
\label{consistentanom}
\end{equation}
The symbol ``Tr'' denotes the trace over the $U(N)$ generators. 
The right hand side of this equation is called the consistent form of the
gauge anomaly since it satisfies the Wess-Zumino consistency condition,
which reads
\begin{equation}
\delta_{\omega_1}{\cal A}(\A,\omega_2)\;-\;
\delta_{\omega_2}{\cal A}(\A,\omega_1)\;=\;
-i\,{\cal A}(\A,[\omega_1,\omega]).
\label{WZeq}
\end{equation}
That the Wess-Zumino condition  holds is a consequence of the fact that
\begin{displaymath}
\big(\delta_{\omega_1}\delta_{\omega_2}\,-\,\delta_{\omega_2}\delta_{\omega_1}
\big)\W[\A]\,=\,-i\delta_{[\omega_1,\omega_2]}\W[\A],
\end{displaymath}
and that the consistent current satisfies the following equation
\begin{displaymath}
(\cJ_\mu^{\rm consistent})^{i}_{\;j}(x)\,=\,\frac{\delta \W[\A]}{\A^j_{\mu\,i}(x)}.
\end{displaymath}

 The reader should note that eq.~(\ref{consistentanom}) leads to the conclusion that the triangle anomaly cancellation condition reads
now
\begin{displaymath}
{\rm Tr} (\T^a\,\T^b\,\T^c)\;=\;0,
\end{displaymath}
whereas its ordinary (for commutative space-time) is
${\rm Tr} (\T^a\,\{\T^b\,\T^c\})\;=\;0$. The reader is referred to  
refs.~\cite{Pepe, Bono} for further comments on this new anomaly
cancellation condition.  The reader should also note that the anomaly
arises because the Jacobian that gives the change of the fermionic measure 
--see eq.~(\ref{measurevar})-- fails to be trivial when properly regularized.
The reader is referred to ref.~\cite{homotopy} for the  computation of    
supersymmetric version of the consistent anomaly above.

One final technical comment. To  obtain eq.~(\ref{consistentanom}) from
eq.~(\ref{plane}) we have dropped an UV divergent contribution which
has the form $\delta_{\omega}{\cal C}(\A,\Lambda)$. This contribution
can be absorbed into a renormalization of $\W[\A]$ since 
${\cal C}(\A_\mu,\Lambda)$ is a polynomial in the $\star$ product of $\A_\mu$ 
and its derivatives.

\subsection{The covariant form of the anomaly}

The covariant form of the chiral gauge anomaly is obtained by defining the 
path integral in eq.~(\ref{effective}) as follows
\begin{equation}
\int d\psi\, d\bar{\psi}\;e^{-{\rm S}[\A]}\;=\;
\int\prod_n \ da_n\,d\bar{b}_n\;e^{-\sum_n\;\lambda_n\,\bar{b}_n\,a_n};
\label{pathintegralb}
\end{equation}
the real numbers $\lambda_n$, which can be chosen so that $\lambda_n\geq 0$, 
and the Grassman numbers $a_n$ and $\bar{b}_n$ are defined thus
\begin{equation}
\begin{array}{l}
{\ds \Big(i\hat{D}(\A)\Big)^{\dagger}i\hat{D}(\A)\varphi_n\,=\,\lambda^2_n\,
\varphi_n,\quad i\hat{D}(\A) \Big(i\hat{D}(\A)\Big)^{\dagger}\phi_n\,=\,
\lambda^2_n\,\phi_n,}\\[9pt]
{\ds \phi_n\,=\,\frac{1}{\lambda_n}\,i\hat{D}(\A)\varphi_n,\;\;{\rm  if}\;
\lambda_n\neq 0,\;\;\; {\rm and}\;\; i\hat{D}(\A)\varphi_n\,=\,0,\;\;{\rm if}\;
\lambda_n =0},\\[9pt]
{\ds \varphi_n\,=\,\frac{1}{\lambda_n}\,\Big(i\hat{D}(\A)\Big)^{\dagger}\phi_n,\;\;}{\rm if}\;
{\ds\lambda_n\neq 0,}\;\;\; {\rm and}\;\;{\ds \;\Big(i\hat{D}(\A)\Big)^{\dagger}\phi_n\,=\,0,\;\;}{\rm if}\;
{\ds\lambda_n =0},\\[9pt]
{\ds \idx\; \varphi_n^{\dagger}(x)\,\varphi_m(x)\;=\;\delta_{nm},\quad
 \idx\; \phi_n^{\dagger}(x)\,\phi_m(x)\;=\;\delta_{nm},}\\[9pt]
{\ds \psi(x)\,=\,\sum_n\,a_n\,\varphi_n(x),\quad
\bar{\psi}(x)\,=\,\sum_n\,\bar{b}_n\,\phi_n^{\dagger}(x)}.
\end{array}
\label{definitions}
\end{equation}
Upon Grassman integration, eqs.~(\ref{effective}) and (\ref{pathintegralb}) 
lead to 
\begin{equation}
\Z[\A]\;=\;\prod_{n}\,\lambda_n\,=\,{\mbox ``}{\rm det}\,
\sqrt{\Big(i\hat{D}(\A)\Big)^{\dagger}i\hat{D}(\A)}\,{\mbox '}{\mbox '}.
\label{covdet}
\end{equation}
Unlike in the consistent anomaly case, $\Z[\A]$ is formally gauge invariant
under the infinitesimal gauge transformation, call it $g$, of $\A_\mu$ in 
eq.~(\ref{gaugetrans}). 
Indeed, $(1-i\omega{\rm P}_{+})\Big(i\hat{D}(\A^g)\Big)^{\dagger}i\hat{D}(\A^g)
 (1+i\omega{\rm P}_{+})=\Big(i\hat{D}(\A)\Big)^{\dagger}i\hat{D}(\A)+
0(\omega^2)$, so that $\lambda_n(\A^g)=\lambda_n(\A)+O(\omega^2)$. 
Pauli-Villars regularization, or zeta function regularization, renders 
rigorous the statements above on the gauge invariance of the  
regularized $\Z[\A]$.  
We thus conclude, as it is the case for commutative space-time, that 
$\delta_{\omega}\W[\A]=0$ for $\W[\A]=-\ln\,\Z[\A]$, with $\Z[\A]$ as defined
by the renormalized counterpart of eq.~(\ref{covdet}). Hence, this $\W[\A]$ 
carries no gauge anomaly. Unfortunately,  the Quantum Action Principle 
\cite{QAP} does not hold for the field theory defined by it. Indeed, now
\begin{equation}
\frac{\delta \W[\A]}{\delta\A^j_{\mu\,i}(x)}\,\neq\, 
\langle \big(\psi^i_{\beta}\star\bar{\psi}_{\alpha\,j}\big)(x)
\big(\gamma_{\mu}{\rm P}_{+}\big)_{\alpha\beta}\rangle\,\equiv\,
\big(\cJ_\mu^{\rm covariant}\big)^{i}_{\;\,j}(x),
\label{horror}
\end{equation}
since it can be shown~\cite{covform} that the covariant current 
$\cJ^{\rm covariant}_\mu$ is not covariantly conserved; i.e. 
\begin{displaymath}
\idx\,\omega^i_{\;j}(x)\, 
\big(D_{\mu}[\A]\,\cJ_\mu^{\rm covariant}\big)^{j}_{\;\,i}(x)\,=\,
{\cal A}^{\rm covariant}(\A,\omega),
\end{displaymath}
with
\begin{equation} 
{\cal A}^{\rm covariant}(\A,\omega)\,=\,
{i\over 32\pi^2}\,{\rm Tr}\int d^4 x\,\omega(x)\,
\varepsilon_{\mu_1\mu_2\mu_3\mu_4}\,\F_{\mu_1\mu_2}\star\F_{\mu_3\mu_4}(x).
\label{covaranom}
\end{equation}
``Tr'' stands for the trace over the $U(N)$ generators.
Let us remark that $\cJ^{\rm covariant}_\mu$ in eq.~(\ref{horror}) is
defined so that $\langle\cdots\rangle$ is given by eq.~(\ref{VEV}) 
with the path integrals defined as in eqs.~(\ref{pathintegralb}) -- 
(\ref{covdet}). It can be shown~\cite{covform} that 
${\cal A}^{\rm covariant}(\A,\omega)$ comes from the nontrivial regularized 
Jacobian that yields the change of the fermionic measure under the chiral 
gauge transformations of eq.~(\ref{gaugetrans}). Here, the regularized Jacobian
is defined by using a regularization function constructed with 
the eigenvalues $\lambda^2_n$ in eq.~(\ref{definitions}).

eq.~(\ref{covaranom}) is  the noncommutative counterpart of the ordinary
result obtained in~\cite{Fuji}. It is clear that ${\cal A}^{\rm covariant}(\A,\omega)$ transforms covariantly under gauge transformations of $\A_{\mu}$ and
that it does not satisfies the Wess-Zumino consistency condition in 
eq.~(\ref{WZeq}). It is not difficult to show by explicit computation that 
the consistent anomaly can be turn into the covariant 
anomaly  by the following redefinition of the currents:
\begin{displaymath}
\big(\cJ^{\rm covariant}_\mu\big)^i_{\;\,j}\;=\; 
\big(\cJ^{\rm consistent}_\mu\big)^i_{\;\,j}\,+\, {\cal X}^i_{\mu\,j}, 
\end{displaymath}
with
\begin{displaymath}
{\cal X}^i_{\mu\,j}\;=\;\frac{i}{48\pi^2} \varepsilon_{\mu\nu\rho\sigma}
\big(\A_\nu\star\F_{\rho\sigma}\,+\,\F_{\rho\sigma}\star\A_\nu\,+\,i\,
\A_\nu\star\A_\rho\star\A_\sigma\big)^i_{\;\,j}.
\end{displaymath} 

\section{The gauge anomaly and the triangle Feynman diagrams}

The contribution to the consistent anomaly equation  --the Euclidean version
of this equation is in eq.~(\ref{consistentanom})-- that comes from the
three point contribution -the famous triangle diagrams-- to $\W[\A]$ reads
thus
\begin{equation}
\begin{array}{l}
{\ds p_3^{\mu_3}\,\W_{\mu_1\mu_2\mu_3}^{a_1 a_2 a_3}(p_1,p_2)^{\eps}=
\,-{1 \over 24\pi^2}\,\varepsilon_{\mu_1\mu_2\alpha\beta}\, 
p_1^{\alpha}p_2^{\beta}}\\[9pt]
{\ds \bigg({\rm Tr}\,\{\Taone,\Tatwo\}\,\Tathree\cos \teta(p_1,p_2)
-i{\rm Tr}\,[\Taone,\Tatwo]\Tathree\sin\teta(p_1,p_2)\bigg).}
\end{array}
\label{trianglecont}
\end{equation}
$\W_{\mu_1\mu_2\mu_3}^{a_1 a_2 a_3}(p_1,p_2)^{\eps}$ denotes the three point
contribution to $\W[\A]$ that involves the Levi-Civita pseudotensor. Notice 
that now we are back in noncommutative Minkowski space-time. As in ordinary
Minkowski space,  one can understand the occurrence of the
gauge anomaly either as an UV effect or as an IR phenomenon,  by evaluating 
in terms of triangle diagrams the left hand side of eq.~(\ref{trianglecont}). 
We shall do this in what remains of the current section; further details can 
be found in ref.~\cite{UVIRorigin}. 

Chiral anomalies have been computed  using Feynman diagrams in refs~
\cite{Iranians} as well.

\subsection{The UV origin of the gauge anomaly}

In dimensional regularization --see~\cite{UVIRorigin}--the left hand side
of eq.~(\ref{trianglecont}) reads
\begin{displaymath}
\begin{array}{l}
{\ds p_3^{\mu_3}\,\W_{\mu_1\mu_2\mu_3}^{a_1 a_2 a_3}(p_1,p_2)^{\eps}=
e^{-\iteta(p_1,p_2)}\,{\rm Tr}\,\Taone\Tatwo\Tathree
\,\triangle_{\mu_1\mu_2}(p_1,p_2;d)\,+}\\[9pt]
\hphantom{p_3^{\mu_3}\,\W_{\mu_1\mu_2\mu_3}^{a_1 a_2 a_3}(p_1,p_2)^{\eps}=~}
{\ds e^{\iteta(p_1,p_2)}\,{\rm Tr}\,\Tatwo\Taone\Tathree
\,\triangle_{\mu_2\mu_1}(p_2,p_1;d),}
\end{array}
\end{displaymath}
where
\begin{displaymath}
\triangle_{\mu_1\mu_2}(p_1,p_2;d)\,=\,
-\,\int{d^d q\over(2\pi)^d}
{{\rm tr}^{\,\eps}\big\{ (\qslash+\poneslash)\,\gambar_{\mu_1}{\rm P}_{+}\,
\qslash\,\gambar_{\mu_2}{\rm P}_{+}\,(\qslash-\ptwoslash)(\bar{\poneslash}+
\bar{\ptwoslash}){\rm P}_{+}\big\}
\over (q^2+i0^+) ((q+p_1)^2+i0^+) ((q-p_2)^2+i0^+)}.
\end{displaymath}
The previous Feynman integral, which is the ordinary one, yields upon 
integration an UV pole at $d-4$. This UV pole is canceled by the 
contribution $(d-4)\varepsilon_{\mu_1\mu_2\alpha\beta}\,p_1^{\alpha}
p_2^{\beta}$ coming from the trace over the gamma matrices.  
eq.~(\ref{trianglecont}) is thus obtained along with the UV interpretation
of it. Of course, this is the same interpretation as in ordinary Minkowski
space-time; an interpretation which can be found in ref.~\cite{Adler}.

\subsection{The IR origin of the gauge anomaly}

To interpret the nonabelian chiral anomaly  on noncommutative Minkowski as an 
IR phenomenon, we shall follow 
Coleman and Grossman~\cite{IR} and compute 
$\W_{\mu_1\mu_2\mu_3}^{a_1 a_2 a_3}(p_1,p_2)^{\eps}$ at the point
\begin{displaymath}
p_1^2=p_2^2=p_3^2=-Q^2,\; p_1+p_2+p_3=0.
\end{displaymath}
Some integration~\cite{UVIRorigin} yields
\begin{equation}
\begin{array}{c}
{\ds \W_{\mu_1\mu_2\mu_3}^{a_1 a_2 a_3}(p_1,p_2)^{\eps}\,=\,}\\[9pt]
{\ds {1\over 24\pi^2}\,\bigg({1\over Q^2}\bigg)\,
\bigg({\rm Tr}\,\{\Taone,\Tatwo\}\,\Tathree\cos \teta(p_1,p_2)
-i{\rm Tr}\,[\Taone,\Tatwo]\Tathree\sin\teta(p_1,p_2)\bigg)}\\[9pt]
{\ds\big(\,\varepsilon_{\mu_1\mu_2\alpha\beta}\, 
 p_1^{\alpha}p_2^{\beta}\,p_{3\,\mu_3}+
\varepsilon_{\mu_3\mu_1\alpha\beta}\, 
p_3^{\alpha}p_1^{\beta}\,p_{2\,\mu_2}+\varepsilon_{\mu_2\mu_3\alpha\beta}\, 
p_2^{\alpha}p_3^{\beta}\,p_{1\,\mu_1}\big).}
\end{array}
\label{IRresult}
\end{equation}
Hence, we conclude that 
$\W_{\mu_1\mu_2\mu_3}^{a_1 a_2 a_3}(p_1,p_2;p_1^2=p_2^2=p_3^2=-Q^2)^{\eps}$
has an IR singularity at $Q^2=0$. The contraction of the right hand side of
eq.~(\ref{IRresult}) with $p_3^{\mu_3}$ erases the singularity at $Q^2=0$ and 
yields the triangle anomaly given in eq.~(\ref{trianglecont}). An explanation
of the  IR origin of the gauge anomaly has been obtained thus.

\section*{Acknowledgements}
The author heartily thanks the organizers of the Euroconference ``Brane New
World and Noncommutative Geometry'' at Villa Gualino for having granted him 
the opportunity of presenting this piece of research in such a beautiful and 
pleasant surroundings. The author is indebted with J.M. Gracia-Bond\'{\i}a 
for an  enjoyable collaboration on anomalies. Finally,  the author is 
grateful to CICyT, Spain, for partial financial support through 
grant No. PB98-0842.

\end{document}